\documentclass[fleqn,10pt]{wlscirep}
\usepackage[utf8]{inputenc}
\usepackage[T1]{fontenc}
\newenvironment{ruledtabular}{\begin{center}}{\end{center}}

\usepackage{graphicx}
\usepackage{dcolumn}
\usepackage{bm}
\usepackage{amsmath}
\usepackage{amssymb}
\usepackage{color}

\usepackage{bm,graphicx,hyperref}
\newcommand{\be}{\begin{equation}}
\newcommand{\ee}{\end{equation}}
\newcommand{\bi}{\begin{itemize}}
\newcommand{\ei}{\end{itemize}}

\title{Direct calculation of the ionic mobility in superionic conductors}

\author[1,2,*]{Alexandra Carvalho}
\author[1,2]{Suchit Negi}
\author[1,2,3]{Antonio H. Castro Neto}
\affil[1]{Institute for Functional Intelligent Materials, National University of Singapore, 117544 Singapore}
\affil[2]{Centre for Advanced 2D Materials, National University of Singapore, 117546 Singapore}
\affil[3]{Department of Materials Science Engineering, National University of Singapore, 117575 Singapore}

\affil[*]{carvalho@nus.edu.sg}

\keywords{solid state batteries, mobility, conductivity, DFT}

\begin{abstract}
 \textcolor{black}{We describe an approach based on non-equilibrium molecular dynamics (NEMD) simulations to calculate the ionic mobility of solid ion conductors such as solid electrolytes from first-principles}. The calculations are carried out in finite slabs of the material, where an electric field is applied and the dynamic response of the mobile ions is measured. We compare our results with those obtained from diffusion calculations, under the non-interacting ion approximation, and with experiment. This method is shown to provide good quantitative estimates for the ionic mobilities of two silver conductors, $\alpha$-AgI and $\alpha$-RbAg$_4$I$_5$. In addition to being convenient and numerically robust, this method accounts for ion-ion correlations at a much lower computational cost than exact approaches.
\end{abstract}

\begin{document}

\maketitle

\section*{Introduction}

The successful expansion of renewable energies and the corresponding liberation from the dependence on fossil fuels require the development of both mobile and stationary storage energy solutions. Li-ion batteries, which store electrical energy directly and are light and compact, have been the solution of choice for consumer electronics, hybrid cars, wearables and medical devices\cite{li201830,murugan2019solid}. However, improvements are needed with regard to safety, cost, storage capacity and lifetime.

Batteries are typically composed of a cathode, an anode and an electrolyte. Commercial Li-ion batteries have a separator to prevent the contact between the cathode and the anode, and a liquid electrolyte solution\cite{deng2015li}. The presence of the liquid electrolyte and the possibility of unexpected chemical reactions in the battery leads to safety concerns with regards to swelling and fire hazard.
All-solid-state batteries, in contrast, employ a solid electrolyte instead of a liquid organic electrolyte. For this reason, they are more stable and need less safety components, leaving room for increasing the packing density\cite{varzi2016challenges,murugan2019solid,fergus2010ceramic,zheng2018review,wu2019utmost}.

The main constraint is the lower ionic conductivity ($\sigma$)  of solid electrolytes,  compared with liquid electrolytes.
The ionic conductivity is a characteristic of the material, and can be directly related to the ionic mobility \textcolor{black}{ ($\mu$), $\sigma=nq\mu$, where $n$ is the density of ions that carry charge and $q$ is the ion charge.}
 A theoretical description of ion drag in solids relating a higher $\mu$ to fundamental properties such as lower density and stiffness has recently been proposed~\cite{rodin2022microscopic}. 
Experimental development have also lead to the discovery of new solid electrolyte materials with ionic mobilities rivaling those of liquids, including garnets, perovskites and NASICON-based materials \cite{wu2019utmost}.

In practice, there are diverse ionic conduction mechanisms, some not yet completely understood.
One of the most studied archetype superionic conductors is $\alpha$-AgI. This is a `molten' sublattice  type superionic conductor, where the charge conducting ions, making up a disordered sublattice, are able to diffuse in a liquid-like fashion, while the remaining atoms display crystalline order\cite{geisel1981nature}. This $\alpha$ phase is stable at ambient pressure only between 147 $^\circ$C and 555 $^\circ$C\cite{mellander}.  The extraordinarily high ionic conductivity of $\alpha$-AgI and its relatively weak temperature dependence are comparable with those of liquid electrolytes making it a paradigmatic example of a solid electrolyte\cite{hull2002crystal}.

Computational methods have been used both to understand the ion conduction mechanisms and to predict which materials are good ionic conductors.
\textcolor{black}{Notably, molecular dynamics simulations under electric field have been  a source of insight on the ion dynamics,\cite{nakamura1994nuclear,futera2020possibility,rozhkov2021effect}, on interface morphology and behaviour,\cite{galvez2020solid,galvez2019ab} and have been used to estimate the ionic conductivity.\cite{das2022structural} Nevertheless, 
the ionic conductivity is instead often estimated based on equilibrium molecular dynamics calculations, using the Nernst-Einstein relationship\cite{zhu2018ab}, which was originally established for gases, and fails when the ionic motion is correlated\cite{pang2021mechanical,marcolongo2017ionic,france2019correlations}.}
Calculations of the mobility based on an exact relationship between the conductivity and particle-particle velocity correlation functions, obtained from linear response theory\cite{kubo1957statistical,van2001first,marcolongo2017ionic,marcolongo2017ionic}, are notoriously difficult to evaluate, requiring large time sampling\cite{france2019correlations} and careful correction for the drift of the origin of coordinates in calculations with periodic boundary conditions\cite{marcolongo2017ionic}.
Since taking advantage of correlations may be a way to surpass the current conductivity limits, it is desirable to be able to preform calculations of the conductivity that include correlations and are at the same time simple and numerically robust.

In this article, \textcolor{black}{we revisit the use of non-equilibrium molecular dynamics simulations in superionic conductors\cite{nakamura1994nuclear,futera2020possibility,rozhkov2021effect,das2022structural} and show} that the ionic mobility, the main figure of merit  for solid electrolytes, can be directly calculated from \textcolor{black}{first-principles}, modelling the drift under an electric field, rather than the diffusion in the absence of the electric field. 

\section*{Results}

We have performed a direct calculation for two silver superionic conductors, $\alpha$-AgI and $\alpha$-RbAg$_4$I$_5$. 

\begin{figure}
    \centering
    \includegraphics[width=12cm]{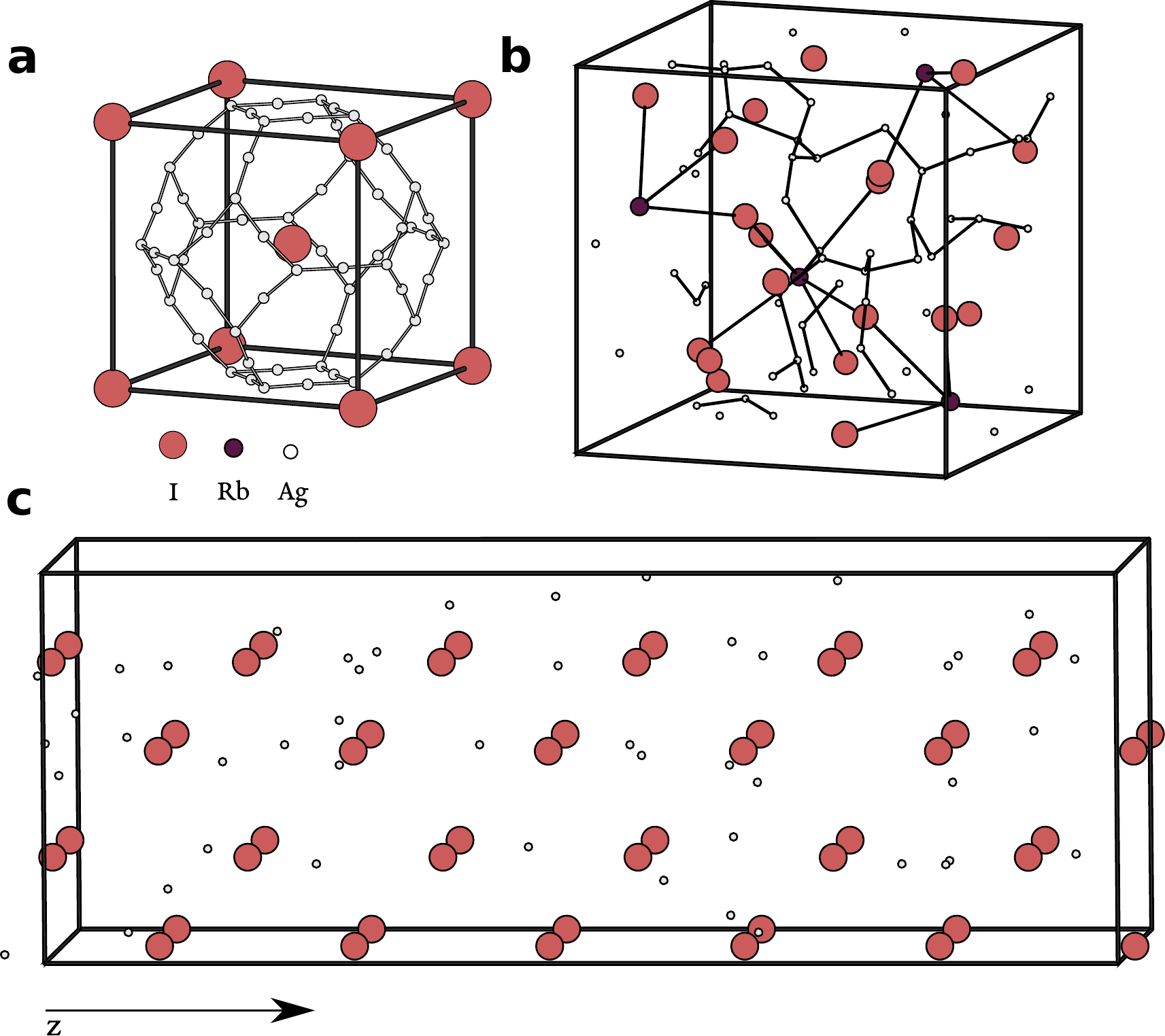}
    \caption{Structure of Ag$^+$ superionic conductors: 
    (a) $\alpha$-AgI, and (b) $\alpha$-RbAg$_4$I$_5$, showing all available positions for Ag. 
    (c) Example of 2$\times$2$\times$6 slab used for calculations.}
    \label{fig:slabs}
\end{figure}

Both the $\alpha$-AgI and the $\alpha$-RbAg$_4$I$_5$ structures have relatively less mobile I or Rb lattices, forming a matrix throughout which the Ag$^+$ ions distribute statistically among a multiplicity of available sites\cite{lawn1964thermal}.
$\alpha$-AgI belongs to the Im$\bar{3}$m space group,
with two I atoms per unit cell forming a body-centered cubic lattice, and the two Ag atoms distributed over the 36 available sites, 24$h$ and 12$d$, with probabilities of 0.07 and 0.027, respectively (Fig.~\ref{fig:slabs}a)\cite{lawn1964thermal}.
The $\alpha$-RbAg$_4$I$_5$ crystal belongs to the P4$_1$32 space group, which has four formula units per primitive cell (Fig.~\ref{fig:slabs}b), with 16 Ag$^+$ ions occupying the available Ag sites, which have been proposed to be 56\cite{hull2002crystal,spencer2013dynamics} or more\cite{burbanoRbAgI,bradley1967relationship,geller1967crystal}.
We have adopted the structure obtained by Spencer et al.\cite{spencer2013dynamics}
as a starting point.

Our calculations employ slabs of material of length $\sim$30-50~\AA\ along the direction of the electric field ($z$), and periodic along the  perpendicular directions (Figure.~\ref{fig:slabs}). We will start by examining in detail the behaviour of the 30~\AA\ slab, which is sufficient to obtain quantitative predictions.
More details can be found under the Methods section.

Our direct calculation of the mobility under the application of a constant electric field resembles the experimental Transient Ionic Current (TIC) technique, which employs DC (direct current) voltage across an electrolyte connected to two blocking electrodes\cite{agrawal1999dc}.
Thus, in respect to boundary conditions, we maintain the atoms at the surfaces of the slab fixed, which is equivalent to having perfectly blocking electrodes, preventing mobile ions from escaping to the vacuum spacing. Calculations where all atoms are free give identical results, for small voltages, provided that we correct for the arbitrary translation of the centre of mass due to the periodic boundary conditions.

Some DFT codes apply by default a slab dipole correction, which creates vanishing internal electric field conditions. Such correction was designed to model ferroelectrics in short-circuit conditions\cite{meyer2001ab}. 
However, in the case of the electrolyte, the internal electric field is not vanishing (except for a fully discharged battery). Thus, a slab dipole correction should not be applied for this particular purpose. The external electric field in the TIC experiments is simply determined by the external DC source, and the internal electric field by the polarisation response of the material.

\subsection*{Electric field response}

\begin{figure}
    \centering
    \includegraphics[width=12cm]{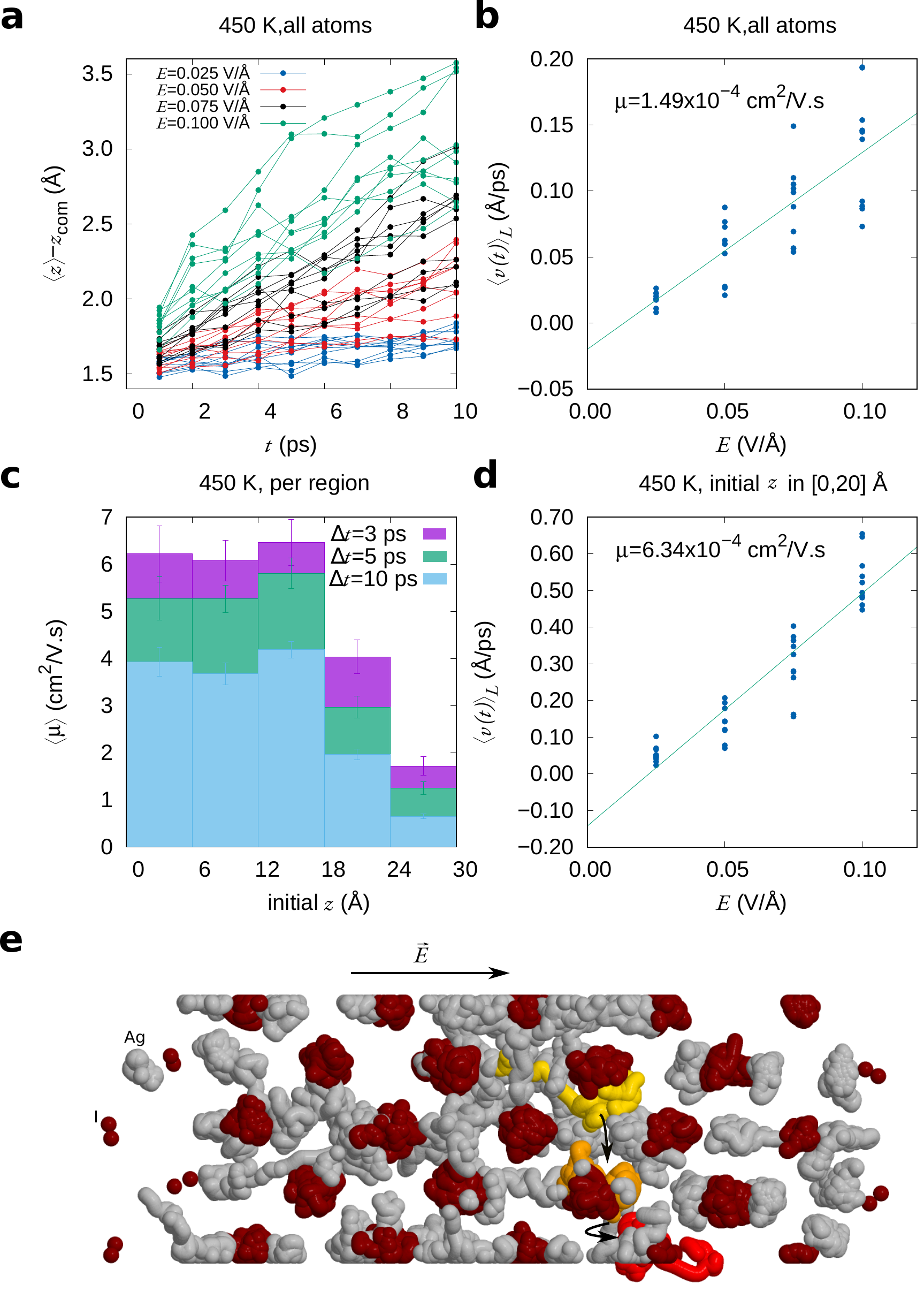}
    \caption{Calculation of the ionic conductivity of AgI at 450~K.
    (a) linear dependence of $z$ on $t$, (b) calculation of the mobility including all atoms in the slab; (c) dependence of the mobility on the initial position of the atoms in the slab, and (d) calculation of the mobility, excluding the atoms at the end of the slab. (e) Illustration of a cooperative jump observed for drift under an 
    electric field of 0.075 V/\AA. The Ag$^+$ ions involved in mechanism
    are highlighted in yellow, orange and red (each one being previously knocked-on by the previous, in this order).}
    \label{fig:linear}
\end{figure}

We are interested in the linear response to the electric field, where conductivity is defined as $\vec{j}=\sigma \vec{E}=nq\vec{v}$, with $\vec{j}$ is the current density, $\vec{E}$ is the electric field, and $\vec{v}$ is the velocity of the ions, and $n$ is the density of ions carrying charge, which we assume for simplicity to be of only one type.

The external electric field $E_0$ is turned on at $t=0$ in a slab that previously was in thermal equilibrium. Thus, for $t\sim 0$, the depolarising field can be neglected.
To keep to a definition  that is consistent to what is measured experimentally, we take $E$ to be the external electric field $E_0$. 
\textcolor{black}{The largest electric field that can be applied in the simulation can be estimated using the expression  $eE_{\rm max}=E_g(L_z)/L_z$, where $E_g(L_z)$ and $L_z$ are the bandgap and the length of the slab, respectively, and $e$ is the electron charge. The bandgap decreases with the slab length due to effect of electron confinement.}
Under application of an external electric field, the coordinates of the Ag$^+$ ions change linearly with time during the first 10~ps (Fig.~\ref{fig:linear}a). Exceptions are (i) for very small times ($<$1~ps), where transport exhibits characteristics of the ballistic regime\cite{vivas2021structural}, and (ii) for high fields close to $E_{\rm max}$, where electronic transitions may already take place due to the presence of defect states.
The calculated velocities, obtained by linear regression, are in turn proportional the electric field (Fig.~\ref{fig:linear}b). 

Some approximations that are implicit to this simulation design are adopting the nominal charge of $+e$ for Ag$^+$, which may be underestimated\cite{wood2006dynamical}, and  neglecting charge transport by I$^-$ ions, which may also lead to an underestimation of the conductivity. Both approximations have been commonly employed by previous studies of $\alpha$-AgI.

\subsubsection*{Finite length correction}
Due to the finite length of the model slab, atoms near the positive end ($z=L$) will have a short distance to run than atoms nearer to the negative end ($z=0$). Hence, if  $L-z_0^i<E\Delta t$, where $z_0^i$ is the initial coordinate of atom $i$ $\Delta t$ is the simulation time, the mobility will be underestimated. Additionally, the interaction with clamped atoms affects the mobility of Ag$^+$ ions.
As can be seen in Fig.~\ref{fig:linear}c, the calculated mobility is approximately constant in the first three-fifths of the slab, but it decays rapidly afterwards. Thus, a more accurate value for the mobility can be obtained if the atoms with $z_0^i>18$~\AA\ are disregarded. The corresponding results are shown in Fig.~\ref{fig:linear}d. The calculated mobility, 6.34$\times10^{-4}$ cm$^2$/Vs is very close to the experimental value (Table~\ref{tab:temp}).

To check for the convergence with respect to the slab length, we have performed additional calculations in slabs consisting of  2$\times$2$\times$10 unit cells. In this case, we have obtained $\mu$=5.18$\pm$1.19\textcolor{black}{$\times10^{-4}$}~cm$^2$/V.s for $z_0^i<20$~\AA\ and $\mu$=8.50$\pm$0.89~\textcolor{black}{$\times10^{-4}$}cm$^2$/V.s for $z_0^i<30$~\AA. We thus see that the calculated mobility increases with the slab length, \textcolor{black}{because the interaction of the Ag ions with the fixed atoms at the $z=0$ and $z=L$ ends decreases, in average. The final value, $\mu$=8.50$\pm$0.89~\textcolor{black}{$\times10^{-4}$}cm$^2$/V.s, converged within $\sim$2$\times10^{-4}$cm$^2$/V.s, is consistent with experiment.}

\subsubsection*{Temperature dependence}

\begin{figure}[h!]
    \centering
    \includegraphics[width=12cm]{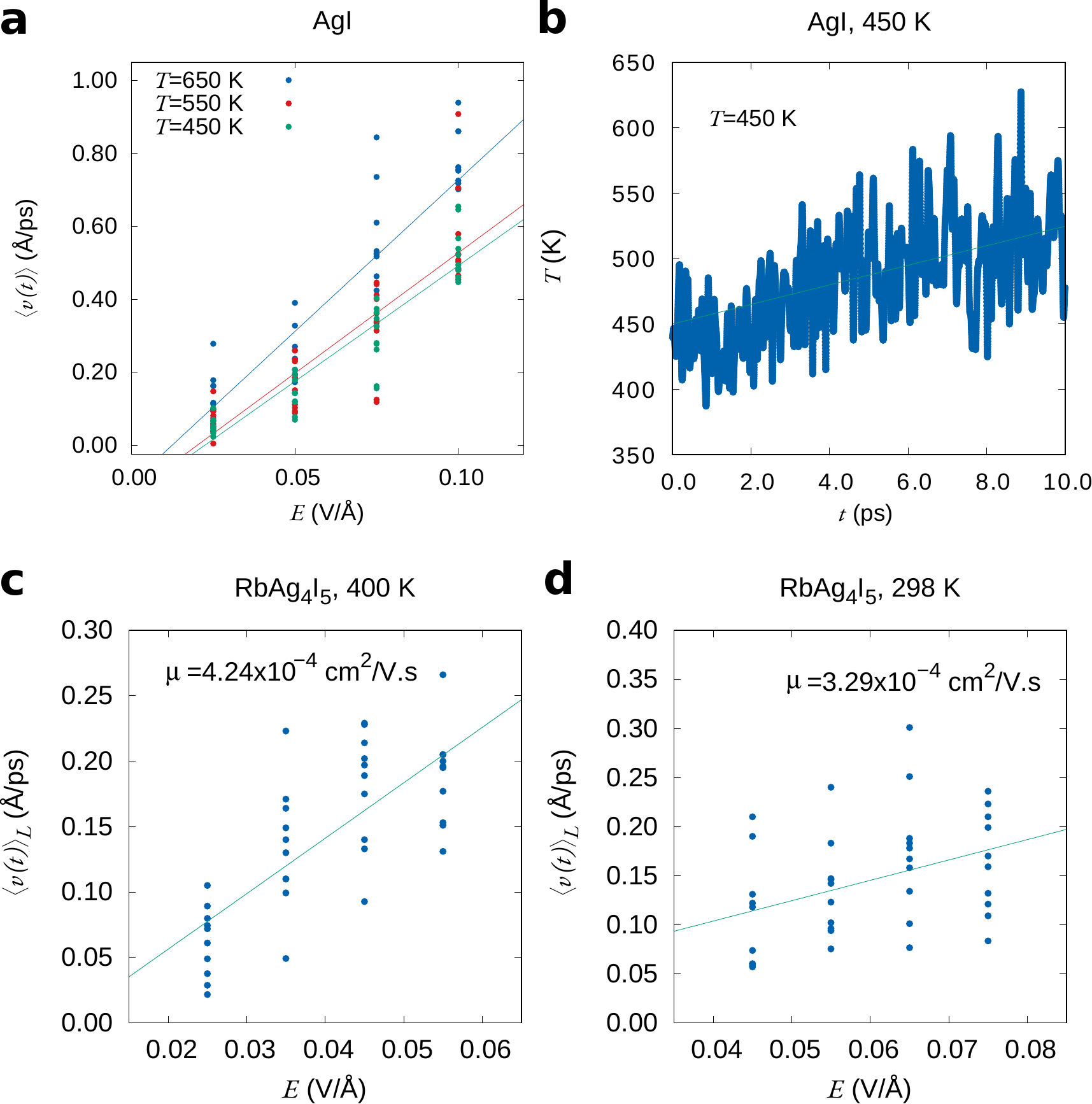}
    \caption{Ionic mobility at different temperatures. For $\alpha$-AgI: (a) dependence of the drift velocity on the electric field and (b) temperature changes during the simulation due to Joule heating, for a starting temperature of 450~K and an electric field of 0.05 V/\AA. The green line is a linear fit. For RbAg$_4$I$_5$ at different temperatures: (a) at 400~K and (b) at 298~K. At 298~K, the ionic mobility of $\sim$3$\times$10$^{-4}$~cm$^2$/V.s is close to the lowest value that we can estimate using this method, with a reasonable number of samples.}
    \label{fig:temp}
\end{figure}

The ionic conductivity of $\alpha$-AgI has a weak temperature dependence\cite{funke1976agi}, due to the small activation energy for Ag$^+$ hopping between sites \textcolor{black}{(see Supplementary Information S2)}. Hall effect measurements also find a weak temperature dependence of the mobility\cite{funke1972ionen}.
We have performed calculations for samples equilibrated to 450, 550 and 650 K. After thermal equilibration using a Nos\'e thermostat,\cite{Nose1984} the dynamical evolution of the system under electric field is calculated using a Verlet algorithm\cite{verlet1967computer}. There is a slight increase of the mobility with temperature (Fig.~\ref{fig:temp}), however not conclusive in comparison with the error bars (Table~\ref{tab:temp}). Above 500~K, $z(t)$ becomes non-linear for the highest electric field (0.1 V/\AA), likely due to the increased defect density and resulting narrowing of the bandgap.
The mobility values presented in Table~\ref{tab:temp} for $T>$450~K are obtained using $E<0.075$~V/\AA.

\begin{table}[h!]
\begin{ruledtabular}
\begin{tabular}{lcccc}
\hline
Material &$T$ (K) & $\mu_{\rm calc.}\times10^{4}$ & $\mu_{\rm exp.}\times10^{4}$ & Ref.\\
         &        &  (cm$^2$/V$\cdot$s) & (cm$^2$/V$\cdot$s) \\
\hline
$\alpha$-AgI
    &  450     &    6.34$\pm$0.40     & 6.36   & \cite{funke1972ionen}\\ 
    &          &                      & {\it 5.8}& \cite{funke1976agi}\\
    &  550     &    6.60$\pm$0.48     & 8.07   & \cite{funke1972ionen}\\
    &          &                      & 10     & \cite{liou1990ionic}\\
    &          &                      & {\it 7.3}    & \cite{funke1976agi}\\
    &  650     &    8.30$\pm$0.98     & 9.79   & \cite{funke1972ionen}\\
    &          &                      & 15-20   & \cite{liou1990ionic}\\
    &          &                      & {\it 8.6} & \cite{funke1976agi}\\
\hline
$\alpha$-RbAg$_4$I$_5$
    &  298     &      3.29$\pm$1.27\\
    &  400     &    4.24$\pm$0.60 & 3.47 & \cite{stuhrmann2002ionic}\\
    &          &                  & {\it 3.3} & \cite{stuhrmann2002ionic} \\
    \hline
\end{tabular}
\end{ruledtabular}
\caption{ Calculated and experimental ionic mobilities at different temperatures. The experimental values shown were obtained from Hall effect measurements or from the fitting of conductivity experiments (conduction mobilities are in italic).
The mobility was extracted from the conductivity values assuming that all Ag$^+$ participate in the conduction.\label{tab:temp}}
\end{table}

In the case of RbAg$_4$I$_5$ at 400~K, non-linear behaviour starts to be observed for an electric field of 0.055~V/\AA, due to the proximity of the valence and conduction band edges for this value of the field (Fig~\ref{fig:temp}-c)). For smaller electric fields, we obtain a mobility  in excellent agreement with experiment (Table~\ref{tab:temp}).
At 300~K, conduction is slower. For the ensemble of 10 starting points that we use, the error bars of $\langle v(E) \rangle$ are comparable to the velocity increment $\frac{dv}{dE}\Delta E$(Fig~\ref{fig:temp}-d). Still, the calculated mobility is surprisingly close to experiment (Table~\ref{tab:temp}). This example illustrates the limitations of this method for conditions of low mobility, which require the use of larger ensemble and higher integration times. Nevertheless, an advantage of this method is that it allows a quick and unambiguous identification of low mobility materials and/or conditions that are of no technological interest.

The lattice constant changes with temperature were neglected in our calculations here\cite{perrott1968heat}. 
Additionally, one of the reasons why it is difficult to determine the temperature dependencee of the mobility is that 
the temperature itself is changing during the application of the electric field,
due to the inelastic collisions of the mobile ions with the matrix and with other mobile ions (Joule effect).

\subsubsection*{Joule effect}

 The heat generation rate by Joule effect is given by $P=VI$, where $V$ is the voltage and $I$ the current, or, in its volumetric form, 
 \be\frac{dP}{dV}=j^2/\sigma=\frac{(nqv)^2}{\mu}.\ee
 The rate of temperature change is then given by
 \be\frac{\Delta T}{\Delta t}=\frac{P}{C_vN},\ee
 where $N$ is the number of moles of material and $C_v$ is the heat capacity at constant volume.
 We observe, as expected, an increase in temperature that is approximately linear in time, as shown in Fig.~\ref{fig:temp}b \textcolor{black}{(see Supplementary Information S3)}. From there we obtain $C_v=169$ J/mol.K, which is higher than the experimental value 80 J/mol.K\cite{perrott1968heat}.

\subsection*{Tracer diffusion}

We now compare the mobility obtained from NEMD to that obtained from equilibrium molecular dynamics simulations, 
using the Einstein relation
\be D = \lim_{t\to\infty} \frac{1}{6t} \langle |\vec{r}(t)-\vec{r}(0)|^2\rangle, \ee
where $\vec{r(t)}$  are the positions of the silver ions, and $\langle ... \rangle$ is the ensemble average. 
This approach assumes that all atoms are independent, 
and the diffusion coefficient $D$ thus defined corresponds experimentally to what is measured by tracer experiments, with very dilute tracers.

\begin{figure}[h!]
    \centering
    \includegraphics[width=8cm]{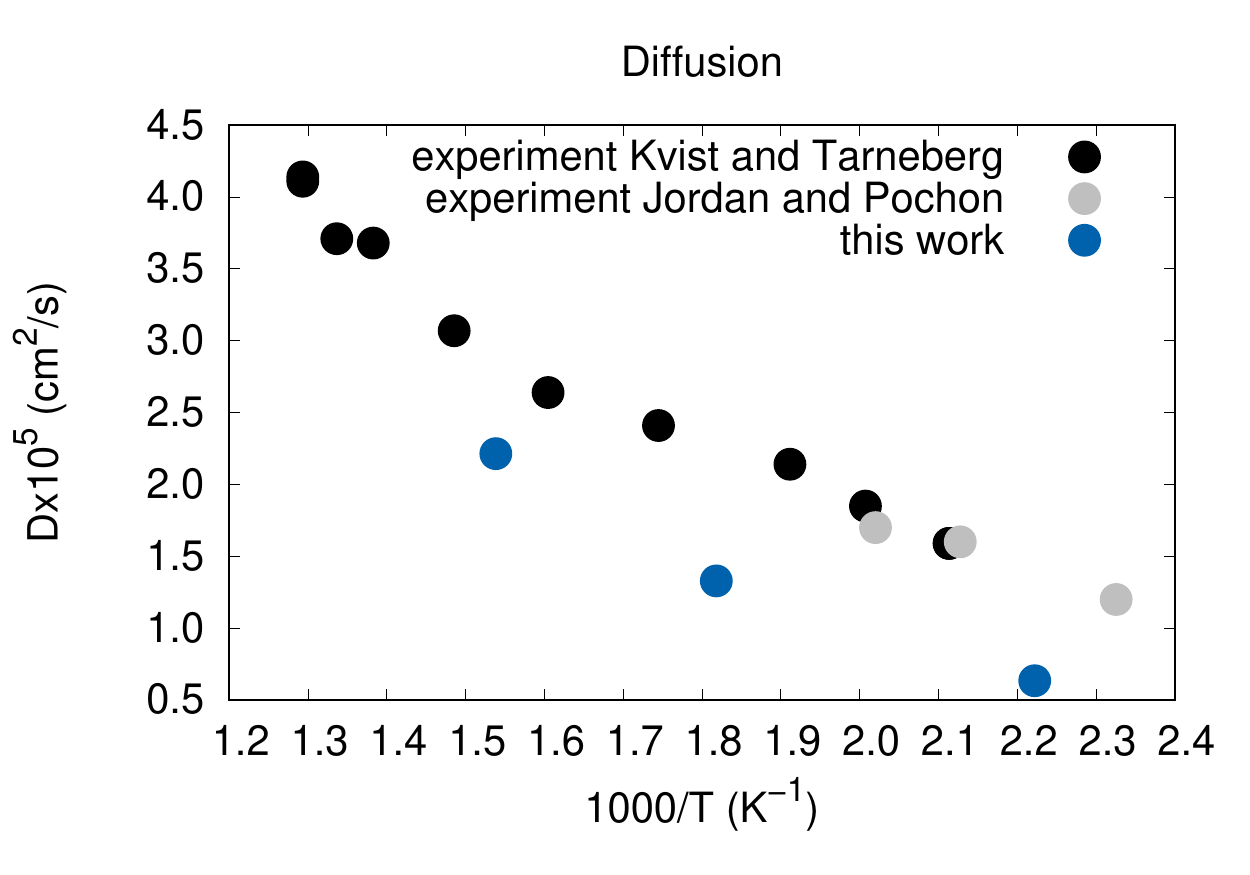}
    \includegraphics[width=8cm]{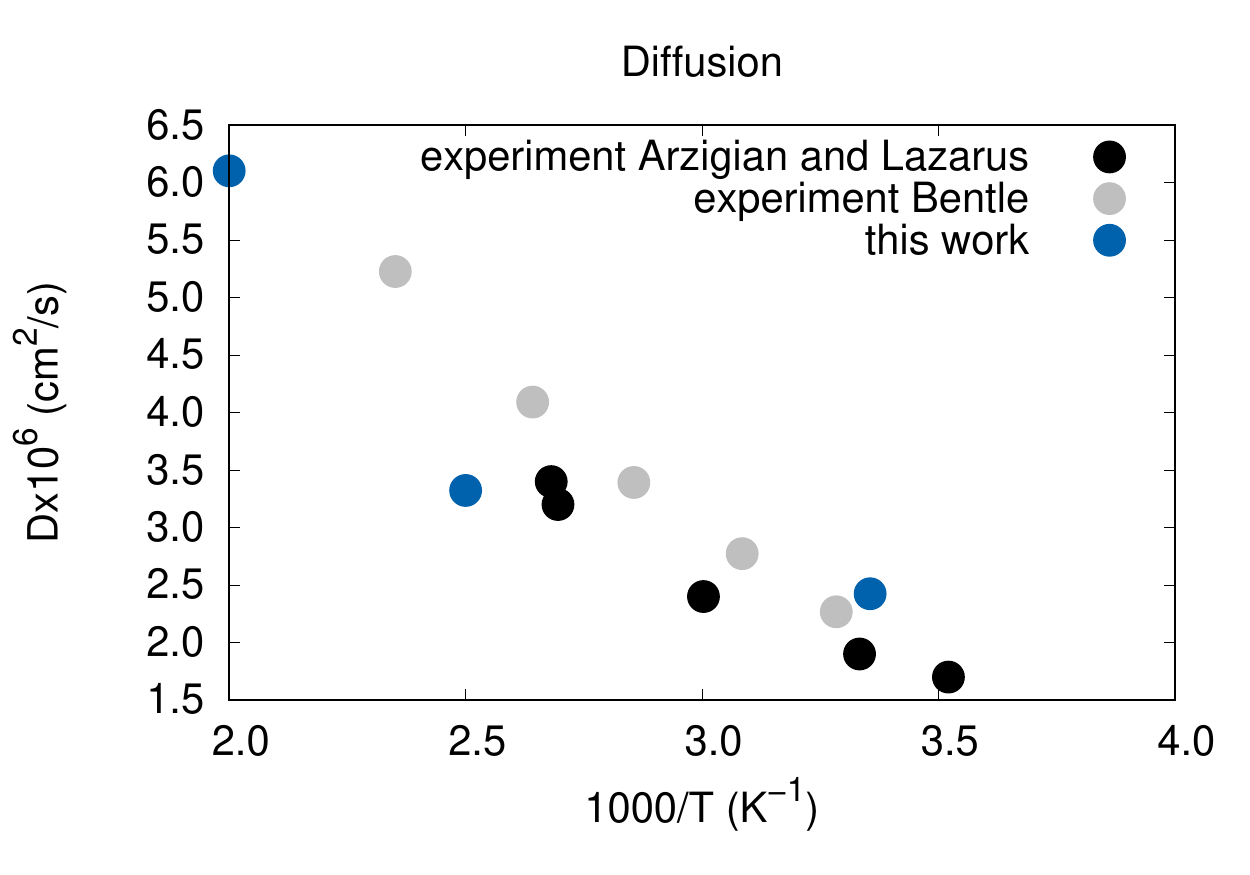}
    \caption{Tracer diffusivities calculated using the Einstein relation. Left: $\alpha$-AgI; Right: $\alpha$-RbAg$_4$I5. The experimental data shown are from Refs.~\cite{kvist1970self,jordan1957m,arzigian1981silver,bentle1968silver}}
    \label{fig:D-T}
\end{figure}

The diffusion coefficient we have obtained for $\alpha$-AgI over 50 ps of diffusion is in good agreement with experiment, showing the same temperature dependence, though systematically underestimated (Fig.~\ref{fig:D-T}). For RbAg$_4$I$_5$, the diffusion constant, estimated for a time of 30-60~ps of diffusion, is also in good agreement with experiment.

Assuming that the atomic displacements are independent and random, the diffusion coefficient can be related to the conductivity via the Nernst-Einstein relation
\textcolor{black}{\be D_{\rm NE}=\sigma_{\rm NE}\frac{kT}{nq^2}. \ee}
Conversely, the mobility can be obtained from the diffusion coefficient:
\be \mu_D=\frac{qD}{kT}.\ee
This mobility is not equivalent to the mobility $\mu$ measured in conduction experiments; still, it is often used as an approximation for $\mu$. 
The ratio from the diffusion and conduction mobilities is the Haven ratio, $H=\mu_D/\mu$, which is determined experimentally by comparing the mobility obtained from tracer diffusion or conductivity experiments. Our calculated values are in reasonable agreement, although smaller than the experimental ones~(Table~\ref{H}).

Since conductivity and tracer diffusivity were calculated by different methods, the calculated Haven ratio may suffer from the lack of error cancellation. Still, it is consistently $<1$, which indicates that the motion of Ag$^+$ ions is correlated. The cooperative `caterpillar' mechanism, whereby Ag$^+$ ions knock-on each other successively, either in a straight line or in zig-zag fashion\cite{yokota1966deviation}, has been previously used to justify the ionic conductivity measurements deviating from the non-interacting Einstein model.
Such knock-on events can easily be observed in our drift trajectories, like the one shown in Fig.~\ref{fig:linear}e.

\begin{table}
\begin{ruledtabular}
\begin{tabular}{lcccc}
\hline
Material & $T$ (K) & $H$ (calc.) & $H$ (exp.) &Ref.\\
\hline
$\alpha-$AgI & 450 & 0.29 & 0.66 & \cite{okazaki1983computer} $^a$\\
             & 550 &      & 0.63 & \cite{okazaki1983computer} $^a$\\
             & 650 &      & 0.60 & \cite{okazaki1983computer} $^a$\\
\hline
$\alpha-$RbAg$_4$I$_5$ & 400 & 0.23 & 0.32
&\cite{arzigian1981silver}$^a$\\
                       & 300 &  & 0.5 & \cite{arzigian1981silver}\\
\hline                  
\end{tabular}
\end{ruledtabular}
\caption{ \label{H} Haven ratio.\\
$^a$ experimental values were interpolated. }
\end{table}

\subsection*{Discussion}

The ionic mobilities calculated for $\alpha$-AgI and $\alpha$-RbAg$_4$I$_5$ by the NEMD method agree well with experimental values reported by both conductivity and Hall effect measurements.
However, there are some considerations that we have to be aware of when comparing the simulations with experiments.
Firstly, our simulations model the drift and can be directly compared to conduction experimental setup. In contrast, there is still no theoretical basis to assume that the Hall mobility is equal to the drift mobility in a solid, especially since ionic Hall effect has only been observed for superionic conductors where cooperative effects seem to play an important role\cite{slifkininbook}.
Thus, we note the agreement with Hall effect measurements even though a formal justification for that is missing.

Additionally, since DC measurements are scarce, we compare our calculated mobilities with those obtained from AC (alternate current) measurements,
assuming that the ionic conductivity is independent on the AC frequency  near the frequency range used in the Hall and conductivity measurements, as it is  believed to be the case below 10$^6$~Hz~\cite{funke2002dynamics}, justifying a direct comparison with our DC simulation. 

We found a report of a DC measurement for $\alpha$-AgI\cite{agrawal1994estimation}, where the conductivity at 450~K was found to be a few times lower than those reported by AC measurements.
Addtionally, we note that Ref.~\cite{agrawal1994estimation} uses a different definition of $\mu$ per mobile carrier -- the carriers in the `ion cloud' first arriving to the end of the pellet. Since we averaged the velocity over all silver ions, for a direct comparison we have to renormalise the mobility in that reference to assume the participation of all silver ions, obtaining $\mu=1.3\times10^{-4}$~cm$^2$/V.s, considerably smaller than other reports. A clarification on whether this difference arises from the frequency dependence of the conductivity or other differences in the method (eg. the presence of memory effects) is needed.

Further, the ionic conductivity in general depends on crystallinity\cite{monty2002ionic}.
In the case of $\alpha$-RbAg$_4$I$_5$, the conductivity of $\mu$m-sized polycrystalline samples has been measured to be 3.3~\% higher than that of single-crystals\cite{raleigh1970ionic}.
For $\alpha$-AgI however, above the temperature of transition to $\alpha$-AgI, the conductivity converges to an unique value independently of the sample shape and size of the crystals\cite{guo2006preparation,guo2007agi}.
This  invariance, and the fact that the conductivity of $\alpha$-AgI has been consistently measured since the earliest experiments, makes $\alpha$-AgI an excellent benchmark system to test our simulation method.

\section*{Conclusions}

The NEMD method \textcolor{black}{employed} here is able to quantify the ionic mobility in solids and therefore can be used to screen and evaluate  potential electrolyte materials, using first-principles or classical molecular dynamics simulations. We now discuss some important considerations when chosing a method to compute the ionic mobility.
Firstly, the advantages of the NEMD method we propose here are as follows:
\bi
\item Ion-ion correlations are fully taken into account, in contrast with calculations using the Nernst-Einstein relationship

\item The NEMD method is easy to converge with a small ensemble and moderate simulation times (provided that the mobility is high enough), and it is numerically very robust with regards to the choice of integration timestep, the presence of anomalous starting points for the trajectory, etc. This makes it easier to use than linear response theory or Green-Kubo methods.
\item The NEMD simulations offer more insight into the physics of the response to the electric field, compared to statistical methods based on equilibrium molecular dynamics. For example, deviations from the linear response to the electric field, or interaction with the surface can be readily observed. Cooperative jumps can be observed and how their direction is determined by the electric field is apparent
\ei
However, these are some of the limitations of the current method:
\bi
\item It is difficult to obtain the temperature dependence of the mobility, due to the heating of the slab as a result of the Joule effect (similar to experiments). Although the present calculations have not taken into account the changes of the lattice parameter with temperature, these can readily be introduced using either ab-initio or experimental parameters.
\textcolor{black}{Still, the calculated activation energy is close to experiment for $\alpha$-AgI, but comparatively smaller for $\alpha$-RbAg$_4$I$_5$. It can be reasonably assumed that the temperature dependence would be more pronounced for electrolyte materials with higher activation energy, such as non Ag-based electrolytes\cite{murugan2019solid}}
\item For poor ionic conductors, it is difficult to estimate the mobility due to the long simulation times involved (see RbAg$_4$I$5$ at 300~K as an example).
Fortunately, the solid electrolytes of technological interest have at present mobilities at least two orders of magniture higher\cite{suci2022review}.
Still, poor ionic conductors can immediately be screened using this present method. For the actual evaluation of the mobility across different orders of magnitude it may be better to use a method based on the evaluation of the energy surface.~\cite{rodin2022microscopic}.
\item For a more accurate calculation of the Heaven ratio, Green-Kubo or Einstein methods may be more appropriate, as the difference between the non-interacting and interacting ion diffusivities can be calculated directly
\ei

 Increasing the probability of cooperative jump mechanisms may be one of the ways to engineer solid electrolytes with improved conductivity.
The \textcolor{black}{NEMD method} is a powerful ally in the design of such materials, and bears testimony to the predictive power of first-principles calculations.

\begin{figure}[h!]
    \centering
    \includegraphics[width=14cm]{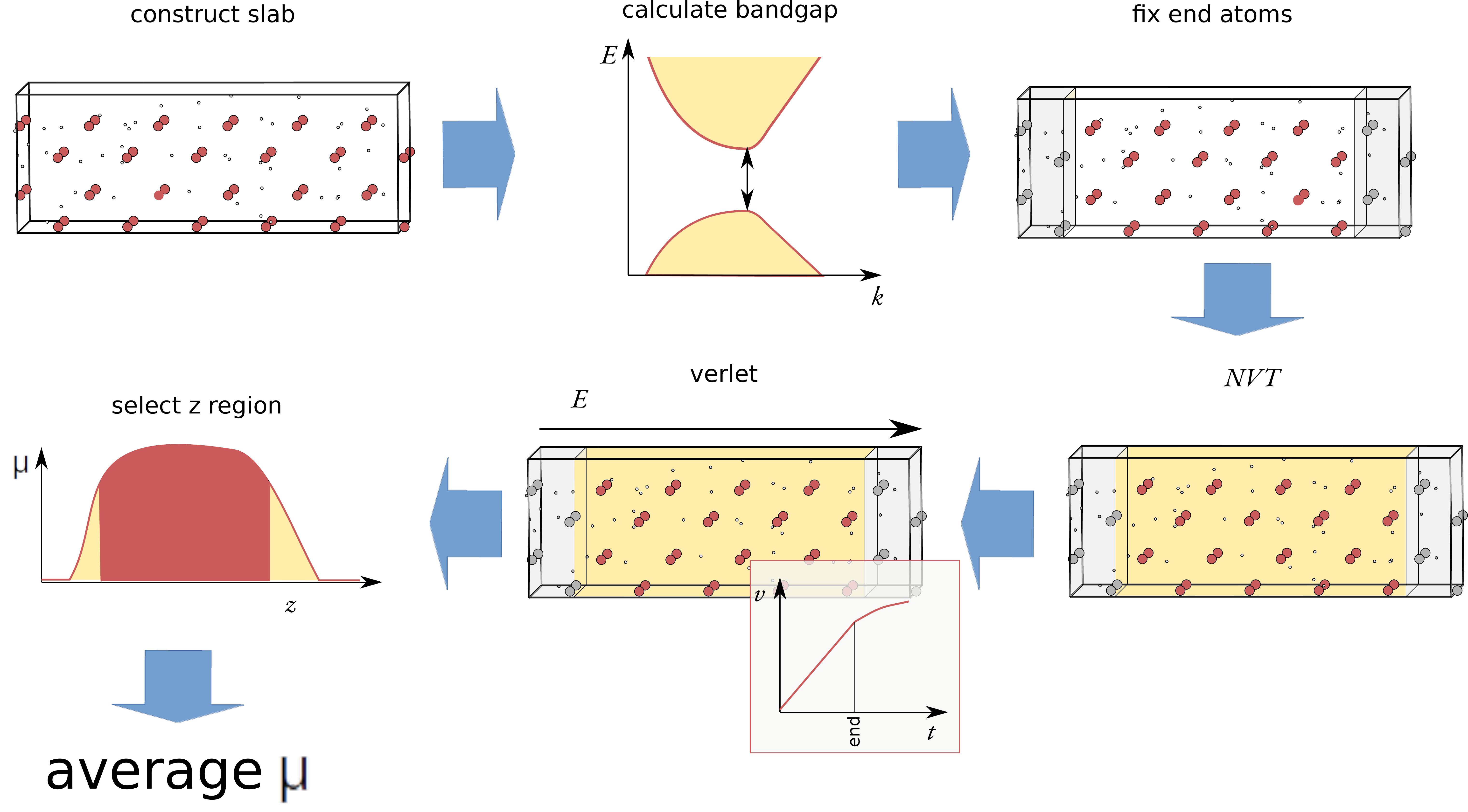}
    \caption{\textcolor{black}{Schematic illustration of the calculation steps (clockwise): construction of the slab; estimation of the maximum electric field from the bandgap; clamping the extremities; equilibration to target temperature; Verlet integration under electric field, and selection of the time interval where $\mu$ is constant; average $\mu$ for the ions that did not start too close to $z_0=L_z$. }}
    \label{fig:method}
\end{figure}

\section*{Methods}

Molecular dynamics simulations were carried out using the SIESTA code~\cite{Soler2002}.
The forces were calculated using the local density approximation (LDA) of density functional theory~\cite{Ceperley1980}, and a Harris functional was used for the first step of the self-consistency cycle. 
The core electrons are represented by pseudopotentials of the Troullier-Martins scheme~\cite{Troullier1991}. 
The basis sets for the Kohn-Sham states are linear combinations of numerical atomic orbitals, of the polarized double-zeta type~\cite{Sanchez-Portal1997,Sanchez-Portal2001}.  
The $\Gamma$-point is used for Brillouin zone sampling. The integration time step used is 1~fs.

\textcolor{black}{
The approach used to calculate the ionic conductivity can be summarised as follows (Figure~\ref{fig:method}):
\begin{enumerate}
    \item The slab was constructed and randomly populated with Ag atoms according to the respective site occupations;
    \item The maximum electric field was estimated from the bandgap and slab length, $eE_{\max} = E_{g}(L_z)/L_z$;
    \item The atoms at both surfaces of the slab were mechanically clamped to prevent drifting atoms from breaking free into the vacuum regions;
    \item The system was equilibrated to the target temperature, in the absence of an electric field, by using a Nos\'{e} thermostat~\cite{Nose1984} (NVT). Different samples/starting points were generated for different equilibration times between 1 ps and 11 ps. The thermostat was turned off at the end of the equilibration ($t=0$)
    \item At $t=0$, the electric field was imposed and the system dynamic equations were integrated using a Verlet algorithm. Then, we selected the time interval for which the ion mobility is constant (linear regime of $v(E)$)
    \item The ion mobility was analyzed as a function of the initial coordinate $z_0$ and average $\mu$ over the region that is not affected by the proximity to $z_0=L_z$ 
\end{enumerate}
 The total integration time for the production runs  was at least 10~ps. The integration time should be long enough to observe ion migration, but short enough so that $E\sim E_{ext}$, where $E_{ext}$ is the external applied electric field (see Supplementary Information S1).
}

The diffusion simulations\textcolor{black}{(equilibrium molecular dynamics)} were carried out using a Nos\'{e} thermostat\cite{Nose1984}, with an equilibration time of at least 5~ps and a simulation time of at least 50~ps. The equilibration time varies between different temperatures and systems and was determined from the mean square displacement, using the Einstein relation. 


\section*{Acknowledgements}
This research / project is supported by the Ministry of Education, Singapore, under its Research Centre of Excellence award to the Institute for Functional Intelligent Materials (I-FIM, project No. EDUNC-33-18-279-V12).
The computational work was supported by the Centre of Advanced 2D Materials, funded by the National Research Foundation, Prime Ministers Office, Singapore, under its Medium-Sized Centre Programme.

\section*{Author Contributions}
AC and SN designed the method and performed the calculations. AHCN defined the problem statement and supervised the project. All authors contributed to the interpretation of results and writing of the manuscript.

\section*{Competing interests}
The author(s) declare no competing interests.

\section*{Data availability}
The datasets generated during and/or analysed during the current study are available from the corresponding author [AC] upon request.


\end{document}